\documentclass{article}

\usepackage{arxiv}

\usepackage[utf8]{inputenc} 
\usepackage[T1]{fontenc}    
\usepackage{hyperref}       
\usepackage{url}            
\usepackage{booktabs}       
\usepackage{amsfonts}       
\usepackage{nicefrac}       
\usepackage{microtype}      
\usepackage{lipsum}
\usepackage{amsmath}
\usepackage{graphicx}
\usepackage{amssymb}
\usepackage{float}

\begin{document}

\begin{flushleft}
{\Large
\textbf\newline{\textbf{Comparing free-space and fibre-coupled detectors for Fabry-Pérot based all-optical photoacoustic tomography}}
}
\newline
\\
Jakub Czuchnowski\textsuperscript{1,\S},
Robert Prevedel\textsuperscript{1,*}
\\
\bigskip
\textsuperscript{1} Cell Biology and Biophysics Unit, European Molecular Biology Laboratory, Heidelberg, Germany
\\
\textsuperscript{\S} Collaboration for joint PhD degree between EMBL and Heidelberg University, Faculty of Biosciences, Germany
\\
\textsuperscript{*}  corresponding author: prevedel@embl.de

\end{flushleft}

\begin{abstract}
All-optical ultrasound detection bears unique advantages for photoacoustics, including wider detection bandwidth, higher signal-to-noise per unit area and lower susceptibility to electromagnetic noise. These benefits have established optical ultrasound detection as a key method for photoacoustic applications in biology and medicine. However, the use of free-space detectors renders this approach sensitive to optical aberrations, which can degrade the pressure sensitivity and result in deteriorated image quality. While spatial mode-filtering through fibers has been proposed to alleviate these problems in Fabry-Pérot based pressure sensors, their real functional advantage has never been properly investigated. In this paper we rigorously and quantitatively compare the performance of free-space and fibre-coupled detectors in a custom correlative setup. We demonstrate the superiority of the latter in terms of both signal level and image quality in realistic all-optical photoacoustic tomography settings.

\end{abstract}

\section{Introduction}

All-optical photoacoustic tomography is an emerging alternative to classical piezoelectric approaches \cite{Wissmeyer:18}. Multiple optical detector types and geometries are constantly being developed and improved with the overarching aim of matching the detection sensitivity of piezoelectric systems. Among them Fabry-P\'erot interferomenter (FPI) sensors are particularly promising, as they combine the ability to measure acoustic waves with high spatial resolution and pressure sensitivity. For this application a pressure sensitive FP device is formed by sandwiching a thin layer (10-100 \textmu m) of elastomere (e.g. Parylene C) between two dichroic mirrors. This optical resonator has then the ability to elastically deform under pressure, modulating the FP interferometer's transfer function (ITF) which is a function of the cavity thickness. The sensor is then interrogated by tuning the laser wavelength to the point of maximum slope on the ITF (so-called bias wavelength) which optically amplifies the incident acoustic wave. In practice, this approach has allowed acoustic sensing in the range of $100-10^6$ Pa with a very broadband frequency response (bandwidth$\sim$0.1-40MHz) \cite{Zhang:08,Jathoul:15} .

As optical devices, FPIs are sensitive to light beam aberrations which can have detrimental effects to their performance under certain conditions. Among different types of optical sensors FP cavities are especially sensitive to aberrations as their sensing principle is dependent on high spatial uniformity of the light beam to facilitate efficient interference \cite{Varu:14}. It was previously shown that both beam and cavity aberrations \cite{czuchnowski2021improving}, surface roughness \cite{marques2021studying} as well as mirror non-parallelism \cite{Varu:14,marques2021analysing} can lead to severe deterioration of the optical sensitivity of the FPI, and that this loss can be partially recovered by the use of aberration correction techniques including Adaptive Optics \cite{czuchnowski2021adaptive}. An alternative method of aberration correction is based on spatial-mode-filtering \cite{czuchnowski2021improving,Marques:20} where higher order spatial modes are removed from the beam with the use of a single-mode fibre coupled decector (FCD). Although FCDs are commonly used in the community \cite{ansari2019use,huynh2016photoacoustic,Buchmann:17} their advantages and disadvantages to free space detectors (FSD) were never directly compared in realistic, experimental conditions. Hence, the real functional advantage of FCDs remained unclear, since important factors such as the effective sensitivity gain as well as expected power losses in FCDs were not previously quantified. In this work we therefore aimed to rigorously compare FCDs and FSDs while taking into account differences in photodetector working points, fibre-coupling efficiency as well as the frequency response of the photodiodes. We found that an FCD is capable of not only significantly improving the optical sensitivity, but also the ultrasound signal level, both of which ultimately translate to improvements in PA image quality. Importantly, FCDs are able to achieve these gains with only moderate losses in the transmitted power as compared to FSDs. Taken all together, these are strong arguments for the use of fibre coupled detectors in FP based photoacoustic sensing and imaging experiments.



\begin{figure}[t]
\includegraphics[width=16.5cm]{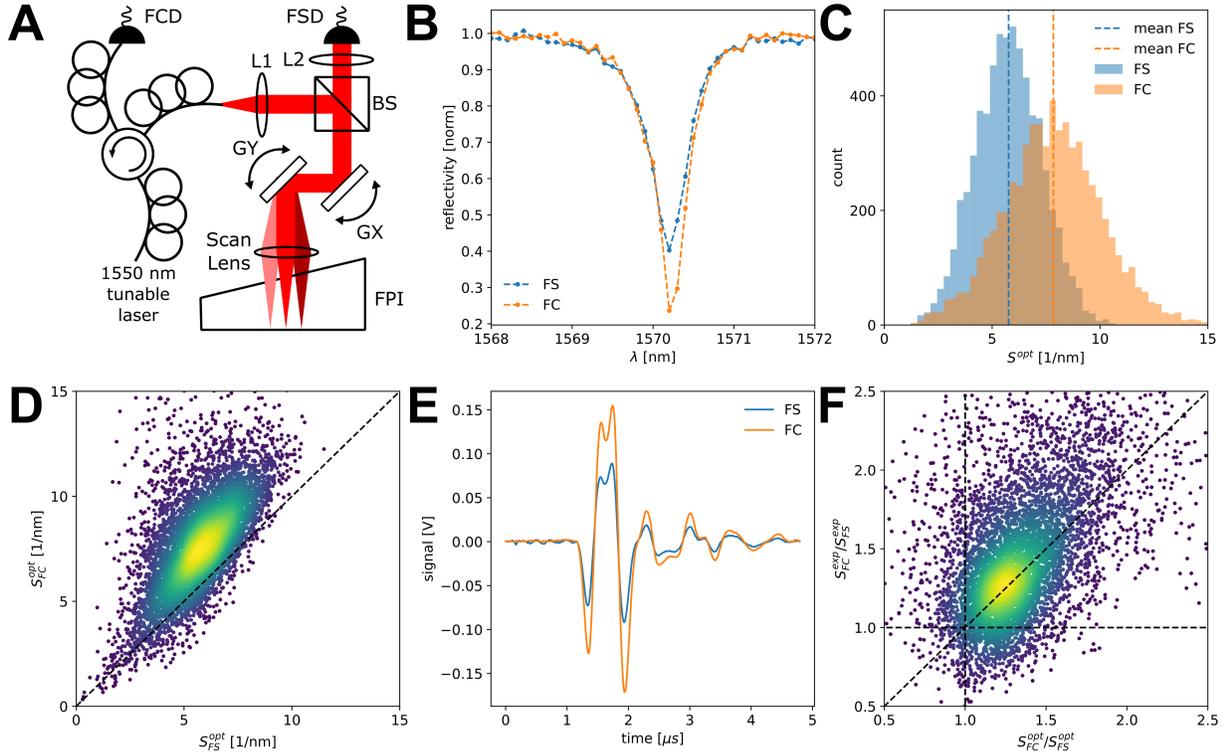}
\centering
\caption{\textbf{A} Schematic of the PAT system for correlative imaging using a free-space and fibre-coupled detectors. \textbf{PD} - photodiode, \textbf{Lx} - lens x, \textbf{GX, GY} - galvo mirros, \textbf{BS} - non-polarising beamsplitter. \textbf{B} Comparison of the free-space and fibre-coupled ITFs showing an increase in visibility in the fibre-coupled condition. \textbf{C} Comparison of the free-space and fibre-coupled normalised optical sensitivity showing an increase in visibility in the fibre-coupled condition.  \textbf{D} Point-by-point comparison of the normalised optical sensitivity showing a point-wise improvement in sensitivity for the majority of spots on the FPI. \textbf{E} Exemplary ultrasound waveform recorded by the free-space and fibre-coupled detectors. \textbf{F} Point-wise comparison between the characterised improvement in optical sensitivity and the measured improvement in ultrasound signal level.}
\label{fig:Fig1}
\end{figure}

\begin{figure}[t]
\includegraphics[width=16.5cm]{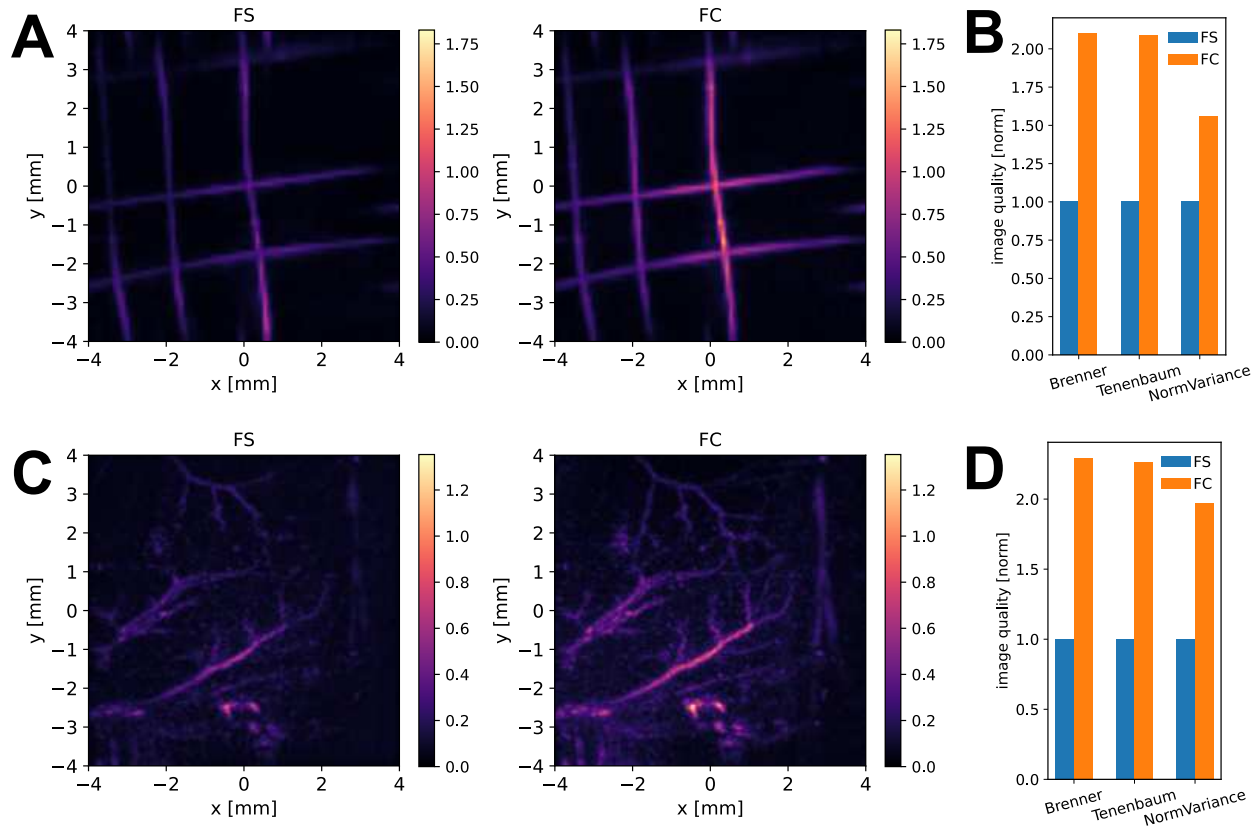}
\centering
\caption{\textbf{A} Experimental PAT images of a wire phantom using fibre-coupled and free-space detectors. Intensity is normalised to the brightest pixel in the free-space image and both images use the same color scale. The images were up-sampled to a 3 times finer spatial grid for better visualisation. \textbf{B/D} Quantification of image quality for \textbf{panel A/D} using common image quality metrics. \textbf{C} Experimental PAT images of a live mouse lower back at 600 nm using fibre-coupled and free-space detectors. \textbf{For panels C and D:} Intensity is normalised to the brightest pixel in the free-space image and both images use the same color scale.}
\label{fig:Fig2}
\end{figure}

\section{Correlative FPI characterisation using free-space and fibre-coupled detectors}

In order to meticulously study the previously suggested \cite{czuchnowski2021improving,Marques:20} differences between free-space and fibre-coupled detectors, we implemented a custom correlative FP based PAT setup (\textbf{Figure \ref{fig:Fig1}A}). We performed simultaneous characterisation of the FPI using both a free-space (FSD) and a fibre-coupled detector (FCD) across $\sim$6500 scan points over 8 x 8 mm sensor surface. We found that the FCD significantly improves the ITF in visibility (\textbf{Figure \ref{fig:Fig1}B}) which is a strong indicator for increased optical sensitivity. We then proceeded to measure the normalised optical sensitivity \cite{czuchnowski2021improving} across the surface of the FPI and observed that the FCD indeed shows an increase in sensitivity (\textbf{Figure \ref{fig:Fig1}C}) which translates to an overall increase on the order of $ \sim 30 \% $ across the whole FPI sensor. To investigate this increase further we analysed the data in a point-wise manner and observe that the increase in sensitivity is uniform with almost all characterised points exhibiting a higher sensitivity with the FCD (\textbf{Figure \ref{fig:Fig1}D}).

To ascertain that the apparent increase in optically measured sensitivity actually translates to improved ultrasound sensing capabilities we performed correlative ultrasound measurements using our system. We observed that the measured ultrasound amplitude is higher by $43\pm 0.11 \%$ (mean $\pm \  2\overline{\sigma}$ for $n\approx4400$ scan positions) for the FCD (\textbf{Figure \ref{fig:Fig1}E}) which is in agreement with the increased optical sensitivity of the FCD. However, it is important to note that differences in the working point between the FCD and FSD need to be taken into account for proper comparison between the conditions as the ultrasound amplitude is directly proportional to the DC level of the photodetector. This effect can be removed by normalising the measured signals by the working point known from the FPI characterisation. We therefore also analysed the data in a point-wise manner by plotting the normalised experimentally measured signal improvement ($S^{exp}_{FC}/S^{exp}_{FS}$) against the increase in normalised optical sensitivity ($S^{opt}_{FC}/S^{opt}_{FS}$) and we observe that the two are in good agreement (\textbf{Figure \ref{fig:Fig1}F}), corroborating the observation that the FCD shows an increase in effective sensitivity. Despite our efforts to accurately quantify both the optical as well as the acoustic gains, there are still off-diagonal outliers present in the data. These presumably stem either from small distortions in the transfer function due to the wavelength dependence of the beam-splitter splitting ratio that affect quantifying the optical sensitivity or from small differences in the frequency response of the photodiodes employed that affect quantifying the ultrasound sensitivity.


\section{Correlative imaging comparing free-space and fibre-coupled detectors}

Having characterised the FPI based system both optically and using ultrasound sources we went on to characterise the photoacoustic imaging properties of the FCD and FSD. We performed photoacoustic imaging of a wire phantom and observed that the reconstructed image intensity is higher for the FCD (\textbf{Figure \ref{fig:Fig2}A}) which is consistent with the characterisation results. Here also the photoacoustic waveforms were normalised to the working point of the photodiode to remove the effect of detector differences from the PA signal amplitude. We quantified the increase in image quality by calculating commonly used image quality metrics and show that for all metrics there is a significant improvement in image quality when using the FCD (\textbf{Figure \ref{fig:Fig2}B}). 

We further compared the performance of FC and FS detectors by performing \textit{in vivo} mouse vasculature imaging experiments. We acquired PAT data from the lower back area using 600 nm excitation to visualise the vasculature in a label-free manner. We observed that also in this case the FCD provides a significantly better image quality (\textbf{Figure \ref{fig:Fig2}C}) which can also be quantified using appropriate metrics (\textbf{Figure \ref{fig:Fig2}D}). This corroborates the superiority of the FCD for both phantom as well as \textit{in vivo} imaging in FPI-based PAT.


\section{Discussion}

We have experimentally demonstrated that mode-filtering with the use of fibre-coupled detectors is capable of significantly improving the sensitivity of FPI based PAT. This finding has important practical implications as for most applications and implementations fibre-coupled detectors are relatively easy to employ as they do not require modifications to the core of the system but only to the peripherally located detector. We note, however, that careful optical design is required because experimentally induced losses in fibre coupling may be significant and potentially disadvantageous over using free space detectors. 

We note that further improvements to the sensitivity can be obtained in principle by combining fiber-based mode filtering with active wavefront modulation approaches (as previously suggested in \cite{czuchnowski2021improving}).
To achieve this, however, several technical challenges need to be overcome in the future, especially on the optical engineering side to increase the beam stability in the system allowing for efficient back-coupling into the fibre in conjunction with active wavefront shaping.


\section{Materials and Methods}

\subsection{Photoacoustic imaging and image reconstruction}

The wire phantom was prepared my suspending a 30 \textmu m nylon surgical suture (NYLON ZO030590) on a custom 3D printed scaffold and submerging in a water bath.

This work was done in accordance to European Communities Council Directive (2010/63/EU) and all procedures described were approved by EMBL’s committee for animal welfare and institutional animal care and use (IACUC). Experiments were performed using C57Bl6/j transgenic mice from EMBL Heidelberg core colonies. An aqueous gel was inserted between the skin and the FPI sensor head to facilitate acoustic coupling. For imaging mice were anesthetized with isoflurane (2\% in oxygen, Harvard Apparatus). Body temperature was kept constant throughout the experiments by the use of a small animal physiological monitoring system (ST2 75-1500, Harvard Apparatus) and eyes were covered with ointment to prevent drying. The diameter of the excitation beam incident on the skin surface was $\approx 1.5$ cm, and the fluence was $\approx 1$ mJ cm\textsuperscript{-2} and was thus below the safe maximum permissible exposure (MPE) for skin \cite{en260825}. 

The FPI used in this study closely resembles previously published FPI designs \cite{Zhang:08} and uses dielectric mirrors with 98\% reflectivity between 1500-1600 nm on a slightly wedged PMMA backing. A $\sim$20 $\mu m$ Parylene C spacer is then deposited in-between the mirrors using vapor deposition. The maximum scan area was 8 × 8 mm\textsuperscript{2} and a typical scan acquired $\sim$ 6000 waveforms each comprising over 1000 time points (sampling rate 125 MHz, ATS9440-128M, AlazarTech). The image acquisition time was $\approx10$ min and was limited by the response time of the interrogation laser. The diameter of the focused interrogation laser beam was 92 \textmu m, which, to a first approximation, defines the acoustic element size. As fibre coupling might cause substantial losses in the transmitted power we compared the transmitted power between the FCD and the FSD and quantified it to be on average $47 \pm 35\%$ (mean \ $\pm \ 2\sigma$, n=6561 point on the FPI). The large variation in the transmitted power probably stems from imperfections of the optical setup (limited telecentricity of the scan lens combined with a non-conjugated galvo system) which could in principle be improved for higher power transmission.

All images shown are based on averaging 3 subsequent excitation pulses at each scan position. Following acquisition of the PA signals, the following protocol was used to reconstruct and display the images: (1) To correct for the effects of photodiode working point the signals were normalised by the power incident on the PD at each scan position which was acquired during the characterisation. (2) For the mouse image the recorded PA signals were interpolated onto a three times finer spatial grid. (3) The tissue sound speed was estimated using an autofocus method \cite{Treeby:11}. (4) A three-dimensional image was then reconstructed from the interpolated PA signals using a time-reversal-based algorithm \cite{Treeby:10} for the mouse image and a back-projection algorithm \cite{Treeby:10b} for the wire phantom with the sound speed obtained in step (4) as an input parameter. The image reconstruction was implemented using k-Wave, an open-source Matlab toolbox (www.k-wave.org; \cite{Treeby:10b}).


\subsection{Quantification of FPI optical sensitivity}

We use an approach based on fitting of the Psuedo-Voigt function:

\begin{equation}
V_{p}(x)=\eta \cdot L(x,f_L)+(1-\eta )\cdot G(x,f_G)
\end{equation}

where $L(x,f_L)$ is a Lorentzian with $f_L$ being the FWHM parameter of the Lorentzian,  $G(x,f_G)$ is a Gaussian with $f_G$ being the FWHM parameter of the Gaussian (see refs. \cite{czuchnowski2021improving,Buchmann:17} for details). We calculate the normalised optical sensitivity from the fit according to our previous work \cite{czuchnowski2021improving}.

\section*{Funding}
This work was supported by the European Molecular Biology Laboratory (EMBL), the Chan Zuckerberg Initiative (Deep Tissue Imaging grant no. 2020-225346), as well as the Deutsche
Forschungsgemeinschaft (DFG, project no. 425902099).

\section*{Acknowledgments}

We acknowledge Florian Mathies, Johannes Zimmermann and Gerardo Hernandez-Sosa from InnovationLab Heidelberg as well as Karl-Phillip Strunk and Jana Zaumseil from Centre for Advanced Materials, Heidelberg University for help with manufacturing of the Fabry-P\'erot interferometers with elastic cavities used in this work. We acknowledge the Mechanical Workshop at EMBL Heidelberg for manufacturing custom opto-mechanical components required for the experimental setup.

\section*{Disclosures}

The authors declare no conflicts of interest.

\bibliographystyle{naturemag}  
\bibliography{biblio}

\newpage
\clearpage

\end{document}